\documentclass[pra,aps,superscriptaddress,showpacs]{revtex4}
\usepackage{graphics}
\begin{document}
  \title{Superfluidity and Anderson localisation for a weakly interacting Bose gas in a  quasiperiodic potential}
\author{Xiaolong Deng}
\affiliation{Universit\'e Joseph Fourier, Laboratoire de Physique et
Mod\'elisation des Mileux Condens\'es, C.N.R.S. B.P. 166, 38042
Grenoble, France}
\author{Roberta Citro}
\affiliation{Department of Physics "E.R. Caianiello'' and
C.N.I.S.M., Universit\`a di Salerno, via S. Allende, 84081
Baronissi, Salerno, Italy} 

\author{Edmond Orignac}
\affiliation{Universit\'e de Lyon, Laboratoire de Physique de
l'\'Ecole Normale Sup\'erieure de Lyon, CNRS UMR5672, 46 All\'ee
d'Italie, 69364 Lyon Cedex 07, France}   
\author{Anna Minguzzi}
\affiliation{Universit\'e Joseph Fourier, Laboratoire de Physique et
Mod\'elisation des Mileux Condens\'es, C.N.R.S. B.P. 166, 38042
Grenoble, France}
\begin{abstract} 
Using exact diagonalisation and Density Matrix
Renormalisation group (DMRG) approach  we analyse the transition
to a localised state of a weakly interacting quasi-1D Bose gas
subjected to a quasiperiodic potential. The analysis is performed
by calculating the superfluid fraction, density profile, momentum distribution
and visibility  for
different periodicities of the second lattice and in the presence
(or not) of a weak repulsive interaction.
It is shown that the transition is sharper towards the
maximally incommensurate ratio between the two lattice
periodicities, and shifted to higher values of the second lattice strength by weak repulsive interactions.
 We also relate our results to recent experiments. %end of abstract
\end{abstract} 
\maketitle
\section{Introduction}

\label{intro}

Ultracold atoms in optical lattices\cite{oberthaler06} represent a
powerful and robust tool for simulating simple quantum systems
with a broad tunability of the parameters, thus serving as
quantum simulators"\cite{feynman82} to reproduce the fundamental
phenomena of condensed matter systems.  One of the most
fascinating and still long-standing problem in condensed matter
physics is represented by the localisation of waves in disordered
media\cite{Anderson}. Actually, evidence of the Anderson
localisation for light waves in disordered media has been provided
by the observation of the modification of the diffusive regime,
indicating a transition from a conductor to an insulator
transition\cite{loc_wave_1,loc_wave_2}. However, a careful analysis is
necessary in order to separate  localisation from absorption
effects\cite{loc_wave_3}.  Anderson localisation in three-dimensions 
has also been recently observed with ultrasound waves \cite{HefeiHu2008}.
In the case of electronic waves in solids,
the situation is made more complicated by the long range coulombic
repulsion between electrons\cite{lee}. Localisation problems can also
be considered in the case of bosonic systems in a disordered medium
such as $^4$He in porous material\cite{Albergamo} or disordered
Josephson Junction arrays\cite{Oudenaarden}. In both of these systems, the
bosons are again strongly interacting. Thus, the
interplay between disorder and interaction
is an important issue in theoretical condensed matter physics.
The combination of ultracold atoms
and optical potentials is offering a novel platform to study
disorder and interaction related phenomena, in which
most of the relevant physical
parameters, including interaction, are tunable\cite{zoller_dis}.
Moreover, a disordered optical potential can be realized either
with laser speckles\cite{inguscio_05}
or with quasi-periodic optical lattices.\cite{inguscio_07}. For both types of disorder, Anderson localisation has been recently experimentally observed.\cite{aspect08,roati08}

In this work we focus on quasiperiodic potentials, which  exhibit
properties common
to both periodic and disordered
systems\cite{diener01}. As in the case of periodic lattices, their
spectrum shows reminiscences of energy bands. On the other hand,
owing to the lack of any translational invariance, they support
the existence of localised states, which behave very similarly to
the ones generated by truly random potentials\cite{grempel82}.
 Localisation of noninteracting particles in quasiperiodic potentials
 has been intensively studied in relation with electronic structures
 of incommensurate solids.\cite{sokoloff85}  The case of a bichromatic
 lattice of Refs.\cite{inguscio_05,inguscio_07} is described by the so
called Harper model\cite{harper} or 1D tight-binding
Aubry-Andr\'e model\cite{aubry,almostMathieu,hofstadter} the
Hamiltonian of which is:
\begin{equation}
H=-t\sum_n( |n\rangle\langle n+1|+|n+1\rangle\langle
n|)+\Delta \sum_n \cos(2\pi n\beta+\phi) |n\rangle\langle n|.
\end{equation}
Here $t$ is the transfer integral between nearest neighbour sites of
the main lattice,
$\Delta$ is the amplitude of the quasiperiodic modulation of the
potential energy,
while $\beta$ is  an irrational number and $\phi$ is a phase. The
Aubry-Andr\'e model can be experimentally realized by
superimposing two optical lattices of incommensurate wavelengths,
 the primary lattice height, $s_1$, being much larger than the
secondary lattice height $s_2$:
\begin{equation}
\label{eq:potential}
V(x)=s_1E_{R1}\cos^2(k_1x)+s_2E_{R2}\cos^2(k_2 x)
\end{equation}
where $k_{1,2}$ is the lattice wavevector and
$E_{R_i}=\hbar^2k_i^2/2m$ the respective recoil energy. At
variance  with the completely random potential which localises all
the states as soon as it is switched on,\cite{Twose} in the
Aubry-Andr\'e case the strength of the potential must exceed the
critical value $\Delta/t=2$ to induce a localised ground state
with a localisation length $1/\ln(\Delta/2t)$ in units of lattice
spacing. Moreover, the localised state at $\Delta/t>2$ is related
to the delocalised state at $\Delta/t<2$ by a duality
transformation. The transition point is self-dual, and its
spectrum is singular continuous, with wavefunctions having a
fractal-like structure\cite{aubry}.

In the recent experiment in Florence by Roati et
al.\cite{roati08}, Anderson localisation has been observed for a
noninteracting $^{39}$K BEC in an incommensurate bichromatic lattice
where the effect of interactions have been cancelled by tuning a
static magnetic field in proximity of a Feshbach resonance to set
the scattering length to zero.  The crossover between extended to
localised states has been studied in detail by looking at the
expansion of the BEC and by studying spatial and momentum
distribution of the atoms, with results in qualitative
agreement with the Aubry-Andr\'e predictions. In particular, they have verified
the scaling behaviour of the critical disorder strength and shown
that while the transition to localisation is expected to be sharp
in the case of a maximally incommensurate ratio
$\beta=(\sqrt{5}-1)/2$, it is broadened for the experimental
parameter $\beta=1.1972$.

So far, the incommensurate lattice supports a localisation transition
 for noninteracting particles, but a richer
physics arises when interactions are present.
A first interesting question is how the
critical disorder strength would be modified in the presence of a
weak interaction. Another question is the nature of the localised
phase of interacting bosons. Indeed, the repulsion between bosons
disfavours the condensation of all particles in the lowest localised
eigenstate, and should instead favour a phases in which although the
bosons are localised, their density is homogeneous.
 This problem has been
discussed in the case of a weakly interacting system in
Ref.\cite{schulte_0506}, where the ground state of the system has
been calculated for a three-colour optical lattice. Introducing
repulsive interactions between the atoms, the numerical
integration of the 1D time dependent Gross-Pitaevskii equation shows that the
ground state wavefunction becomes a superposition of many
single-particle localised states, which add up to form an overall
extended state. Similar results for a bichromatic lattice have
been presented in the work by Lye et al.\cite{inguscio_07},
and Quantum-Monte Carlo calculations for the case of
large interaction strengths
have been performed by Roscilde \cite{Roscilde08a,Roscilde08b}.

Besides the modification of the critical disorder strength,
the combination of lattice and interaction can lead to new phases that
compete with the localised state.
In absence of disorder, bosons on a lattice with repulsive
interactions display, for commensurate filling,
 a superfluid (SF) to Mott insulator (MI) transition as
the repulsion is increased\cite{Fisher}, with the superfluid phase
displaying large density fluctuations and a gapless excitation
spectrum, while the Mott phase is incompressible and has a gap in
the excitation spectrum. If one considers both repulsive
interactions and disorder, these two effects will compete: while
disorder makes the bosons localised, short-range repulsive
interaction energy increases as the square of boson density and
hence the total energy of the system is minimised by depleting the
localised condensate towards a more uniform density distribution.
As a result, in a lattice Bose gas with short-range interactions a
novel Bose-glass (BG) phase, non superfluid yet compressible,
emerges between the superfluid and the
Mott-insulator\cite{Fisher}. For the specific case of a
one-dimensional Bose gas subjected to an uncorrelated disorder (in
absence of a lattice), the phase diagram has been obtained by
Giamarchi and Schulz \cite{GiaSchu}, showing that while for zero
interactions the system is always localised, for nonzero values of
the repulsive interactions a superfluid phase is possible at small
values of disorder. Ref.\cite{GiaSchu} also predicted that  the
non-superfluid (Bose-glass) phase of an interacting Bose gas is
expected to differ markedly from the non-interacting
Anderson-localised (AG) phase, e.g. the density profile of a  Bose
glass phase is rather uniform, in contrast to the highly
inhomogeneous density profile of an ideal Bose gas in a disordered
potential where all the particles occupy the lowest
single-particle localised orbital. The case of lattice hard core
bosons with nearest neighbour repulsion was considered in
\cite{Doty}. It was found that a superfluid dome could be present
for strong enough nearest neighbour attraction and weak enough
disorder. The phase diagram of a disordered, interacting lattice
Bose gas in one-dimension has been the subject of several
numerical investigations by exact diagonalisation \cite{Runge},
quantum Monte Carlo methods \cite{Batrouni,Svistunov}, strong
coupling expansions\cite{monien96}, and Density-Matrix
renormalization group approaches (DMRG)\cite{DMRG_disorder}, that
have established the existence of a Mott insulating phase
separated from the superfluid phase by a Bose glass phase for
disorder not excessively strong. For stronger disorder, these
numerical studies have established that only the Bose glass and
the superfluid is present. Also, the existence of a superfluid
dome in the phase diagram for a
 quasiperiodic lattice potential
has been
obtained in the commensurate case\cite{DMRG_disorder,Roux08,nostro}.

Here we focus on the quasi-1D Bose gas subjected to a
quasiperiodic potential in the presence or absence of weak
interactions and try to analyse the scaling behaviour of the
critical interaction strength by exact diagonalisation or DMRG. In
particular, we characterise the transition by means of the
behaviour of the superfluid density, the density profile and the
visibility. In the case of open boundary conditions, where the
superfluid density is undefined, another physical quantity is also
introduced for the non-interacting case, 
i.e. the inverse participation ratio and its results are discussed.

\section{The model}
\label{sec:model}

\subsection{The Hamiltonian}

We consider a one-dimensional Bose gas at zero temperature
subjected to a bichromatic lattice potential as the one of
Eq.(\ref{eq:potential})

\begin{eqnarray}
\label{eq:cont_model} H=&&\int_{-\infty}^{\infty} dx
\psi_b^\dagger
(x) \left( -\frac{\hbar^2}{2m} \nabla^2 +V(x)\right)\psi_b(x)\nonumber \\
&& +\frac g 2 \int _{-\infty}^{\infty} dx
\psi_b^\dagger(x)\psi_b^\dagger(x)\psi_b(x)\psi_b(x),
\end{eqnarray}
where $\psi_b(x)$ is the bosonic field operator, $m$ is the atomic
mass and $g=4 \pi \hbar^2 a_s /m$ is the contact interaction expressed in terms of the scattering length $a_s$. When the primary lattice
height $s_1$ is large compared to the atomic recoil energy $E_{R,1}$  and to the height of the secondary lattice  $s_2$ , we can map the
Hamiltonian on a Bose-Hubbard model \cite{jaksch}:

\begin{eqnarray}\label{eq:1} H &=&
-t\sum^{N-1}_{i=1} (b^{\dagger}_i b_{i+1}+h.c.)+
\frac{U}{2}\sum^{N}_{i=1} n_i(n_i -1)   \nonumber
\\&-&\mu\sum^{N}_{i=1} n_i  +
\sum^{N}_{i=1}\Delta_i n_i,
\end{eqnarray}
where $b^{\dagger}_i$, $b_i$ are bosonic field operators on the
site $i$, $t$ is the hopping amplitude, $U$ is the on-site
interaction, $\mu$ is the chemical potential, $N$ is the total
number of lattice sites; the parameters $U,t$ are related to those
of the continuum model (\ref{eq:cont_model})(e.g. see Refs.
\cite{jaksch,Buchler03}) and their respective dependence upon the
recoil energy, lattice depth and scattering length can be
calculated both analytically and
numerically\cite{bloch_review_08}. The effect of the second
lattice is to induce a modulation of the on-site energies
according to $\Delta_i=\Delta \cos(2 \pi \beta i+\phi)$, with
$\Delta\propto s_2$ and $\phi$ being a phase shift while in the
experiments $\beta=k_2/k_1$. By taking the parameters of
\cite{inguscio_07_pra} one can estimate $s_2/t=2.5 --- 53.$

In order to characterise the localisation transition, we evaluate
the following observables in the case of periodic boundary
conditions: (i) the superfluid fraction as the response to twisted boundary conditions, $f_s= \frac{m L^2}{N \hbar^2} \frac{\partial^2 E_\theta^N}{\partial \theta^2 }$,
which on the lattice model reads
\begin{equation}
\label{eq:fs} f_{s} = \frac{N^2}{N_P
t\theta^2}\left(E^{N_P}_{\theta}-E^{N_P}_{PBC}\right),
\end{equation}
where $N_P$ is the particle number, $E^{N_P}_{PBC}$ is the ground
state energy and $E^{N_P}_{\theta}$ is the ground state energy with
boundary conditions twisted by an angle $\theta$ \cite{Batrouni},
(ii) the spatial density profile $\langle n_j\rangle= \langle
b^\dagger_j b_j \rangle$, and (iii) the momentum distribution,
given by the Fourier transform of the one-body density matrix
\begin{equation}
n(q) = {\cal N} \sum_{lm}{\rm e}^{iq (l-m) \alpha}\langle
b^{\dagger}_l b_m\rangle.
\end{equation}
with $\alpha=\pi/k_1$ being the primary lattice spacing and $ {\cal N}$
a normalisation constant.

\subsection{Numerical  methods}

The ground state properties of the interacting Bose gas in the
bichromatic lattice have been determined by the Density Matrix
Renormalization Group (DMRG) method\cite{White_DMRG,DMRG_review}
and exact diagonalisation (ED) method (in the noninteracting case).

Concerning DMRG, it is a quasi-exact numerical technique widely
employed for studying strongly correlated systems in low
dimensions. Based on the renormalization, it finds efficiently the
ground state of a relatively large system  with quite high
precision.
We have been considering a system with periodic boundary conditions
and used first the infinite-size algorithm to build the Hamiltonian up
to the length $L$, then we have been resorting to the finite-size
algorithm to increase the precision within many sweeps. Since the
Hilbert space of bosons is infinite, to keep it finite we have been
choosing the maximal number $n_{max}$ of boson states approximately of the order
$5 \langle n \rangle$ , where $ \langle n \rangle$ is the average number of bosons per site,
varying $n_{max}$ between $n_{max}=10$ and
$n_{max}=25$. The number of eigenstates of the reduced density matrix
are chosen in the range $150-250$.  The number of sites in the DMRG is
$N=50$. The initial phase of the quasi-periodic potential is chosen
$\phi=\pi$.

In the noninteracting case, $U=0$, we have been employing the
exact diagonalisation method on a chain of 50 up to 1000 sites.
The superfluid fraction depends on the energy change under
twisted-phase boundary conditions. We checked that the superfluid
fraction computed from (\ref{eq:fs})  does not vary strongly at
critical points $\Delta_c/t$ when small angles (say $\theta=0.1$
rad) are replaced with large angles (say $\theta=\pi$). Thus, the
location of the critical point is not sensitive to the choice of
the twist angle in Eq.~(\ref{eq:fs}). This allows us to use
anti-periodic boundary conditions (i.e. a twist angle
$\theta=\pi$) to calculate the superfluid fraction (both in ED and
DMRG).  We have also analysed the variation of the superfluid
fraction as a function of the shift $\phi$ of the quasi-periodic
potential in the Hamiltonian~(\ref{eq:1}). Under some initial
phases with fixed length of chain, we found that the critical
points are shifted to a value much smaller than $\Delta/t=2$. This
behaviour is associated with the formation of an isolated state
below the continuum of extended states which is localised near the
edges. The origin of this localised state is simply the fact that
the on site potential $\Delta_i$ does not have the periodicity of
$N$, thus causing strong fluctuations of the potential near the
edges in the system with periodic or aperiodic boundary
conditions.  In our calculations we  choose the initial phase
such that no spurious localised state is formed near the edges,
ensuring that the critical points are close to $\Delta_c/t=2$. In
addition, when we perform finite size scaling we  consider a
sequence of periodic approximants to the potential $\Delta_i$
associated with larger and larger ring sizes. In the
noninteracting limit we consider also open boundary conditions
(OBC).  In this case the superfluid density (\ref{eq:fs}) fails to
be a good indicator of the localisation transition, but  the
inverse participation ratio\cite{thouless74} can be employed.

\section{Analysis of the localisation transition}
We proceed here to present the results for the localisation
transition occurring at increasing height of the secondary
lattice,  with the purpose of analysing the localisation
transition at varying the parameter $\beta$, i.e. the periodicity of
the secondary lattice. Although mathematically the transition is
expected {\em only for irrational values} of $\beta$
\cite{almostMathieu}, in the numerical
calculations $\beta$  is a rational number. In the experiment, $\beta$
is fixed by the ratio of the wavelengths of the two laser beams
used for creating the optical lattices, which are only known up to a
finite number of significant digits, so that the rational or
irrational character of $\beta$ cannot be decided. In principle, if
$\beta$ is rational, the system is in a periodic potential, and it thus
 possesses only extended states forming a finite number of
bands for any strength of the potential.  However, if
$\beta$ is  given by the ratio of two mutually prime integers, with a large
denominator, the periodical behaviour can only be observed on  a very
large length scale.
Since, in contrast to the ideal case studied in
Mathematics\cite{almostMathieu}, both in simulations and in
experiments the number of sites of the lattice is finite,
 the very large lengthscale periodicity mentioned above may
not be accessible in practice, and the question arises on the occurrence and
the nature of the Anderson localisation transition (or
crossover). More precisely, in the case of $\Delta \gg t$, if the
periodicity of the potential is $N$, degenerate perturbation theory
provides an estimate of $\sim t(t/\Delta)^{N-1}$ for the bandwidth of the
extended states. On timescales small with respect to the inverse of
the bandwidth, particles will exhibit the appearance of localisation.
We thus investigate here the Anderson localisation behaviour of a
{\em finite lattice} and {\em rational} $\beta$ parameter, which
is relevant for the ongoing experiments.

\subsection{Noninteracting case}
We begin our analysis by considering first the noninteracting limit $U=0$,
 In this case, the ground state of the Hamiltonian Eq.(\ref{eq:1})  can be obtained by
exact diagonalisation.

\subsubsection{The inverse participation ratio}

In the noninteracting regime it is possible to adopt several
definition of localisation.
One which turns out to be  particularly convenient for our problem is the inverse participation ratio in the ground state\cite{thouless74}.
This can be expressed as
\begin{equation}
P_0=\sum_i |\psi_i^{(0)}|^4
\end{equation}
where $\psi_i^{(0)}$ is the normalised ground state wavefunction of the noninteracting Hamiltonian (\ref{eq:1})
 on the basis $|i\rangle$ of the lattice site occupation number.
The inverse participation ratio is related to the probability of
return to the initial point \cite{thouless74} and  allows to
distinguish between extended and localised states: it decays as
$1/N$ where $N$ is the  number of lattice sites in the extended
regime and goes to  a constant value in the localised regime.

\begin{figure}
% if it does not compile replace by .eps
\resizebox{0.8\columnwidth}{!}{
\includegraphics{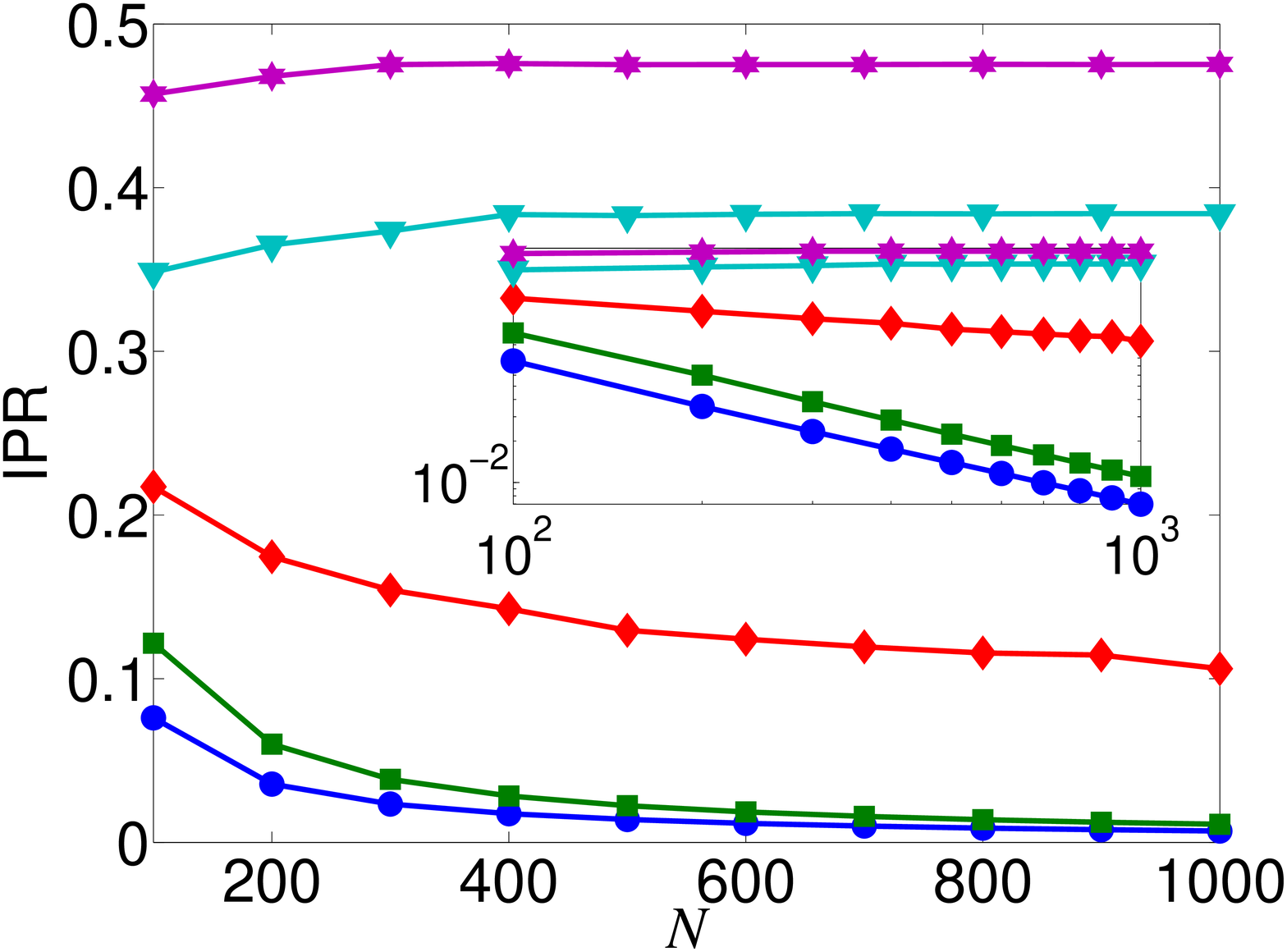}
}
\caption{(Color online) Inverse participation ratio IPR$=1-\delta w/N$ (adimensional)
  for a noninteracting Bose gas in a quasiperiodic potential with
  $\beta=(\sqrt{5}-1)/2$ with open boundary conditions as a function
  of the  number of sites in the chain for various values of the
  height of the second lattice $\Delta$ (in units of $t$), from
  bottom the values are $\Delta/t$=1.9, 1.95, 2, 2.05, and 2.1 .
The inset shows the same quantity in logarithmic scale.}
\label{fig:ipr_n}
\end{figure}

The inverse participation ratio can be extracted from the numerical calculations for a system with open boundary conditions by computing the average number fluctuations,
\begin{equation}
\delta w= \frac{1}{N}  \sum_i \left[ \langle n_i^2\rangle- \langle n_i\rangle^2 \right].
\end{equation}
Indeed, using the expansion of the boson site operators $\hat b_i$ on the operators $\hat b_\eta$ corresponding to the eigenvectors  $\psi_i^{(\eta)}$ of the Hamiltonian, ie $\hat b_i=\sum_\eta \psi_i^{(\eta)}\hat b_\eta$, and the fact that in the ground state all the $N$ particles occupy the state  $\psi_i^{(0)}$  we obtain a relation between the number fluctuations and the inverse participation ratio
\begin{equation}
\delta w=N(1-\sum_i |\psi_i^{(0)}|^4).
\end{equation}
Since the  inverse participation ratio has two distinct large-$N$ behaviour in the extended and localised regimes, it is a good indicator to identify the transition from the extended regime, where $\delta w/N=1+O(1/N)$, to the localised regime where $\delta w/N=1-const$. Fig.\ref{fig:ipr_n} shows the change in behaviour of the inverse participation ratio across the localisation threshold.

\begin{figure}
% if it does not compile replace by .eps
\resizebox{1.0\columnwidth}{!}{
\includegraphics{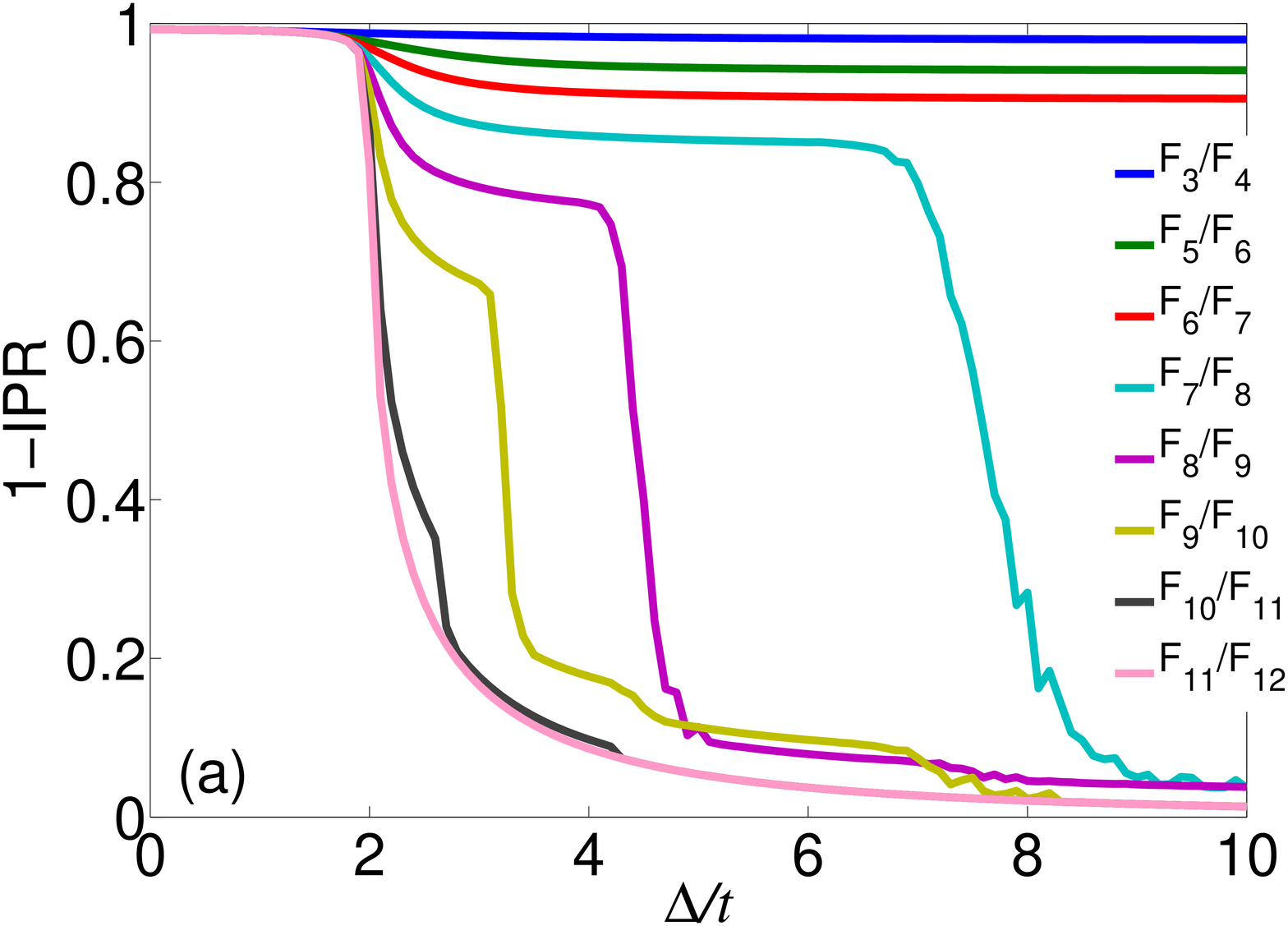}
\includegraphics{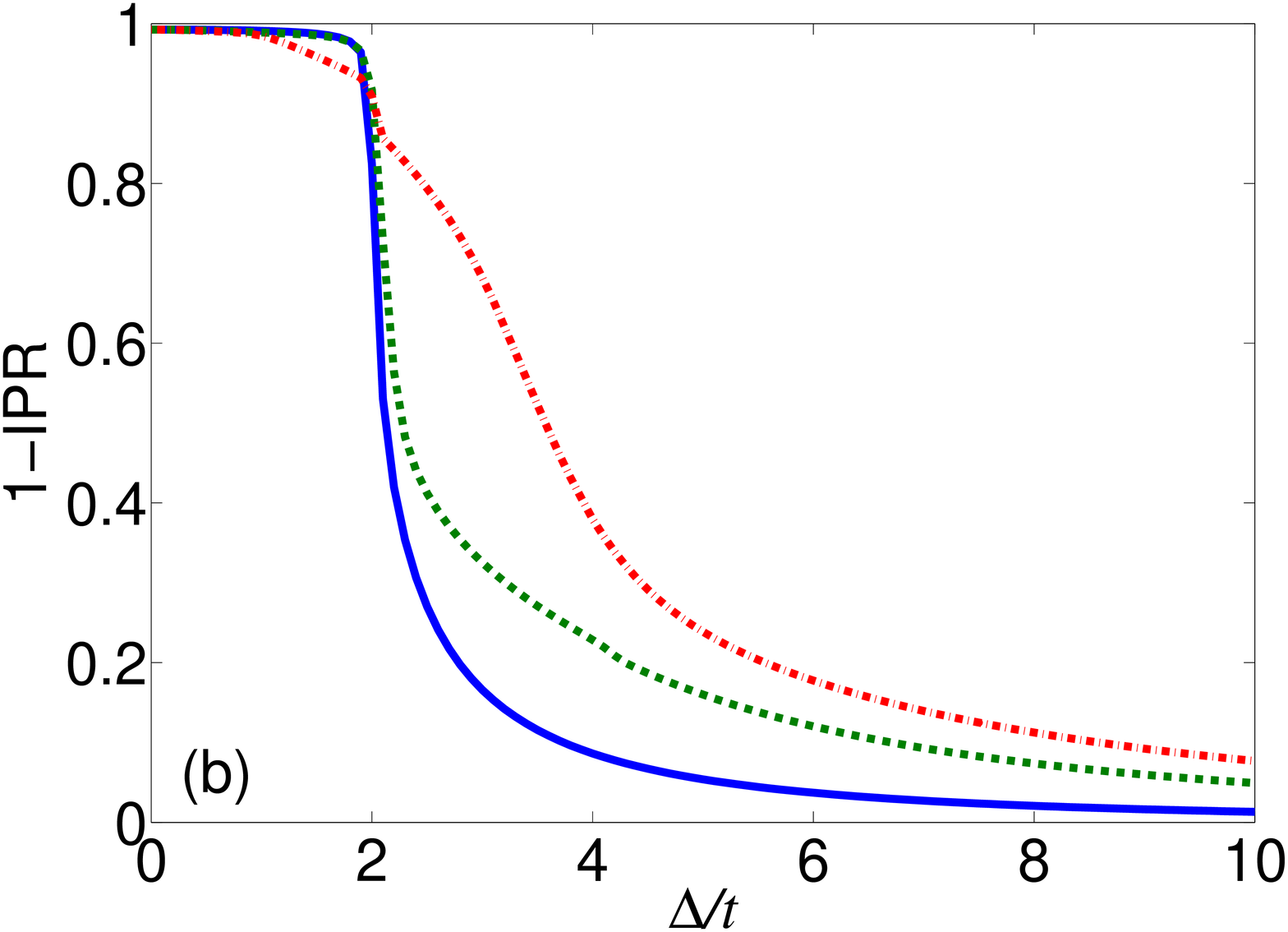}
}
\caption{(Color online) Average number fluctuation $\delta w$ (adimensional) for a noninteracting Bose gas in a quasiperiodic potential with open boundary conditions as a function of the  height of the second lattice $\Delta$ (in units of $t$) for a number of lattice sites $N=200$, at varying choices of
$\beta$.  First panel, subsequent approximants of the number
$\beta=(1+\sqrt{5})/2$ according to the Fibonacci sequence. Second
panel  $\beta=(\sqrt{5}-1)/2$ (blue solid line),  $\beta=830/1076$
(green dotted line), and  $\beta=1032/862$ (red dash-dotted line).}
\label{fig:ipr}
\end{figure}

Figure \ref{fig:ipr} shows the average number fluctuation as a
function of the height of the secondary lattice $\Delta$ for
different choices of the quasiperiodic potential. In the first
panel, we choose for $\beta$ subsequent approximants of the
irrational number $(1+\sqrt{5})/2$, according to the Fibonacci
sequence: $F_1=1$, $F_2=1$, $F_{n+1}=F_n+F_{n-1}$, as  $\lim_{n\to
\infty} F_{n+1}/F_n=(1+\sqrt{5})/2$. The figure shows that the
transition at $\Delta/t=2$ becomes more and more pronounced as the
order of the approximation increases. In the second panel, we show
the behaviour of the inverse participation ratio for three choices
of $\beta$ which we will consider throughout the paper:   the
maximally irrational value $\beta=(\sqrt{5}-1)/2$, the value
adopted in the first Florence experiment \cite{inguscio_07}
$\beta=830/1076$, and the value used in the second Florence
experiment  \cite{roati08} $\beta=1032/862$.

\subsubsection{The superfluid fraction}
\begin{figure}
% if it does not compile replace by .eps
\centerline{\resizebox{0.75\columnwidth}{!}{
\includegraphics{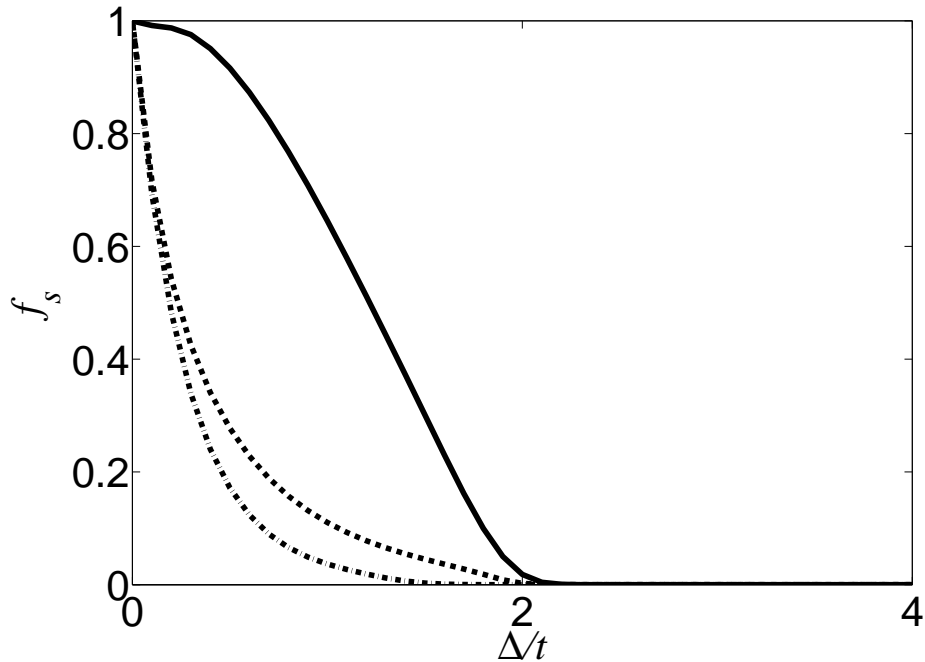}
}}
\caption{Superfluid fraction for a noninteracting Bose gas in a quasiperiodic potential as a function of the height of the  second lattice $\Delta$ in units of the hopping strength $t$ for three choices of the quasiperiodicity period $\beta$:  maximally irrational value $\beta=(\sqrt{5}-1)/2$ (solid lines), first Florence experiment
\cite{inguscio_07}  $\beta=830/1076$ (dashed lines), and second
Florence experiment  \cite{roati08} $\beta=1032/862$ (dot-dashed
lines). All the curves are drawn for $\phi=\pi$ and $N=50$.
}
\label{fig:1}
\end{figure}

Since
the inverse participation ratio is  defined only in
the noninteracting limit, we further
characterise the
transition by monitoring the superfluid fraction, which is nonzero
in the extended phase and vanishes in the localised phase.
We  obtain the superfluid fraction from
the change of the ground state energy of the system to twisted
boundary conditions, Eq.(\ref{eq:fs}). Figure \ref{fig:1} shows
the behaviour of the superfluid fraction as a function of the
height of the secondary lattice $\Delta$, for various choices of
the quasiperiodicity parameter $\beta$ and for a number of lattice
sites equal to 50, which is close to the experimental conditions.
We have considered three possible values of $\beta$:    the
irrational number $\beta=(\sqrt{5}-1)/2$, and  the values
$\beta=830/1076$, $\beta=1032/862$ adopted in the Florence experiments \cite{inguscio_07,roati08}.
From the data we see that the superfluid fraction vanishes around the value
$\Delta/t=2$, as predicted by the Aubry-Andr\'e model
\cite{aubry}, the details however depend on the choice of $\beta$.
 We remark that the
transition is particularly smooth in the case of the second
Florence experiment, where the potential has five minima very
close in energy.

\begin{figure}
% if it does not compile replace by .eps
\resizebox{1.0\columnwidth}{!}{
\includegraphics{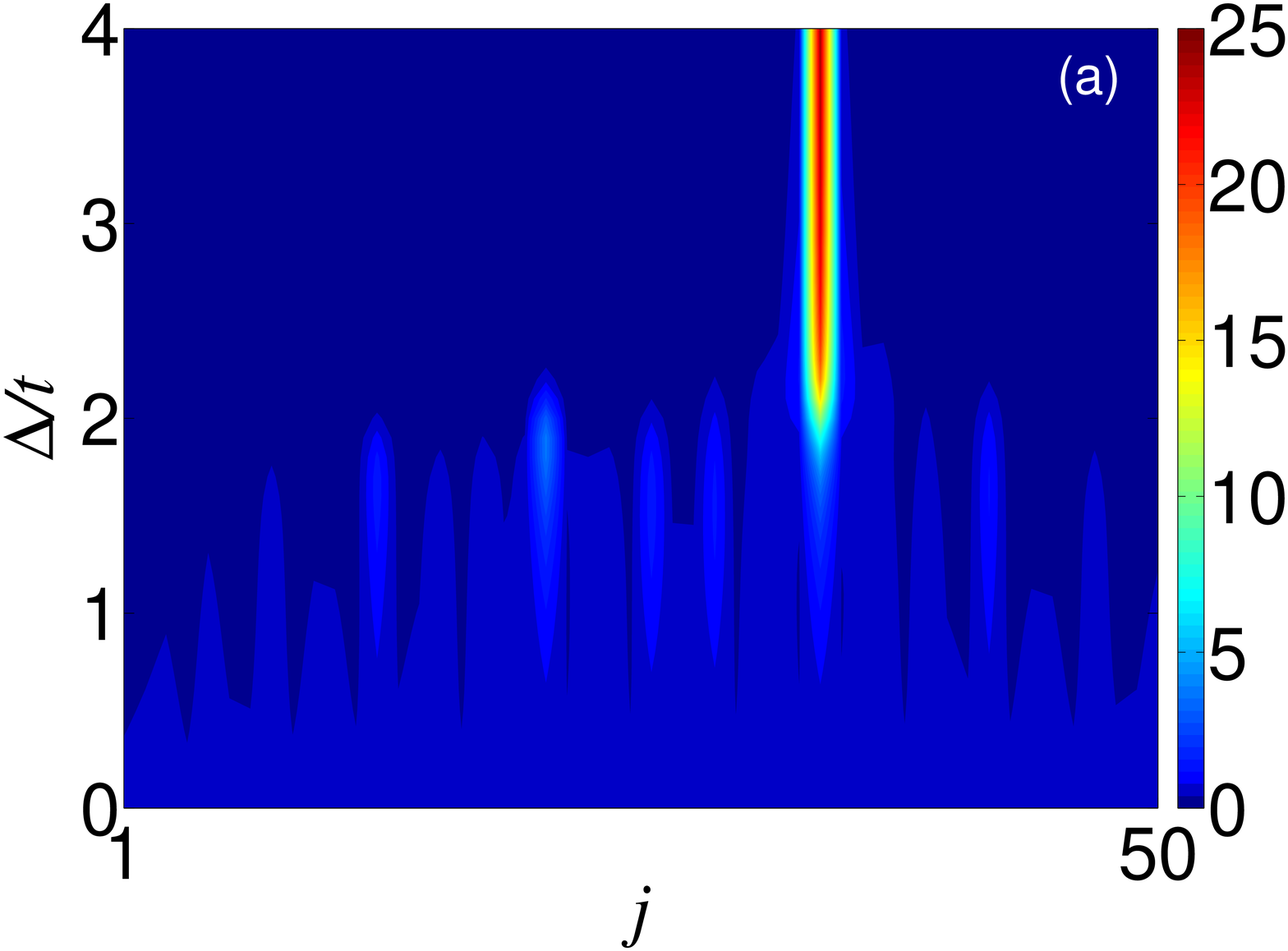}
\includegraphics{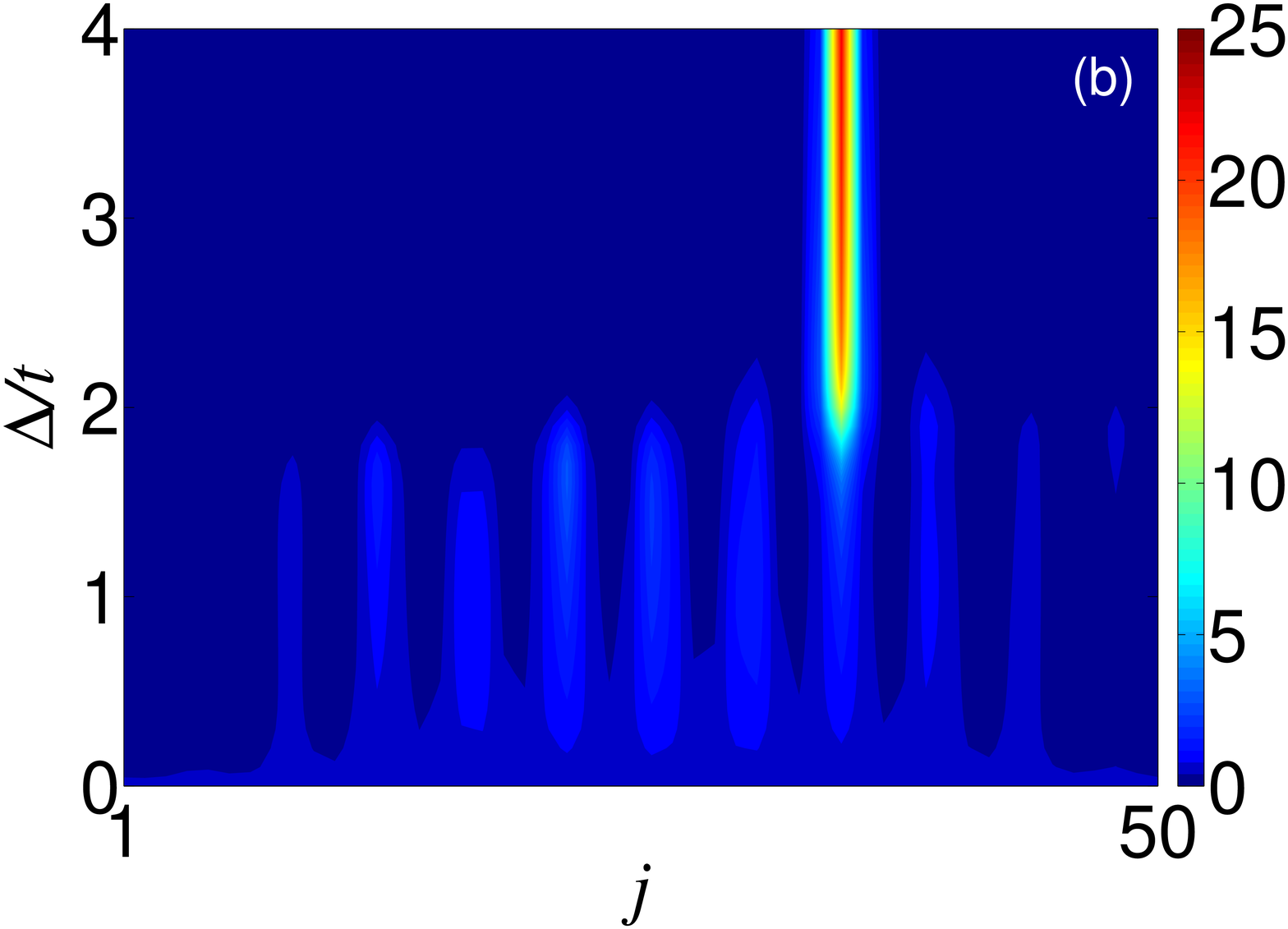}
\includegraphics{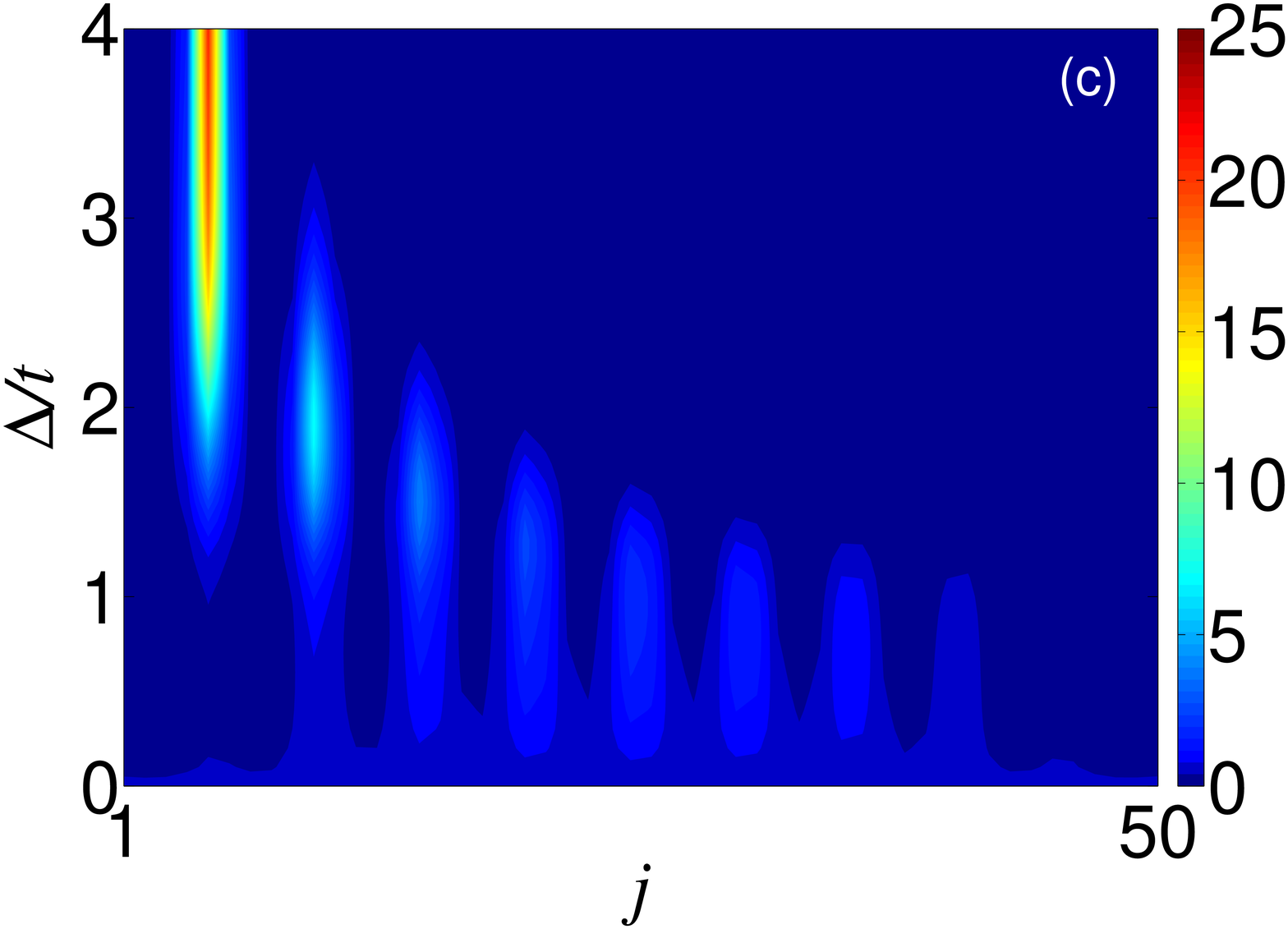}
}
\caption{(Colour online) Contour plot of the density profile for a noninteracting Bose gas in a quasiperiodic potential with periodic boundary conditions at increasing height of the second lattice $\Delta$ for a number of lattice sites $N=50$, at varying choices of
$\beta$:  $\beta=(\sqrt{5}-1)/2$ (a),  $\beta=830/1076$ (b), and  $\beta=1032/862$ (c).In the three cases  localisation is clearly visible for $\Delta>2t$.  The value of the phase shift has been chosen $\phi=3\pi/10$ for (a) and (b), and $\phi=-2\pi/5$ for (c).}
\label{fig:2}
\end{figure}

The Anderson localisation transition can be also seen by looking at
the spatial density profiles, illustrated in Fig. \ref{fig:2} at
increasing values of $\Delta$. In correspondence to the transition the
profile displays a clear change of behaviour as all the particles tend
to cluster around the localised state. The position of the
Anderson-localised peak corresponds to the position of the absolute
minimum of the quasiperiodic potential.

\subsubsection{Finite size scaling}
\label{sec:fss}

All the results of the previous section were obtained for a system of
finite length. In such a system, there is no actual localisation
transition as the localisation length can never exceed the length of
the system and the spectrum always remains a pure point one.
In order to interpolate the results to the case of an
infinite system which possesses a true localisation transition between
localised and extended state, we make a finite-size scaling
Ansatz.\cite{barber}
We thus assume that the superfluid fraction is of the form:
\begin{equation}
  \label{eq:fss-ansatz}
  f_s(N,\Delta)=N^{-a} \Phi\left(\frac N {\xi(\Delta)} \right),
\end{equation}
where $\xi(\Delta)$ is a characteristic length that diverges at the
localisation transition as $|\Delta-\Delta_c|^{-\nu}$. The function
$\Phi(x)$ is assumed to be regular for $x\ll 1$ and to behave as
$\Phi(x)\sim x^{a}$ for $x\gg 1$ on the extended side,
so that the superfluid fraction
becomes independent of system size in the thermodynamic limit.

\begin{figure}
  \resizebox{0.85\columnwidth}{!}{
  \includegraphics{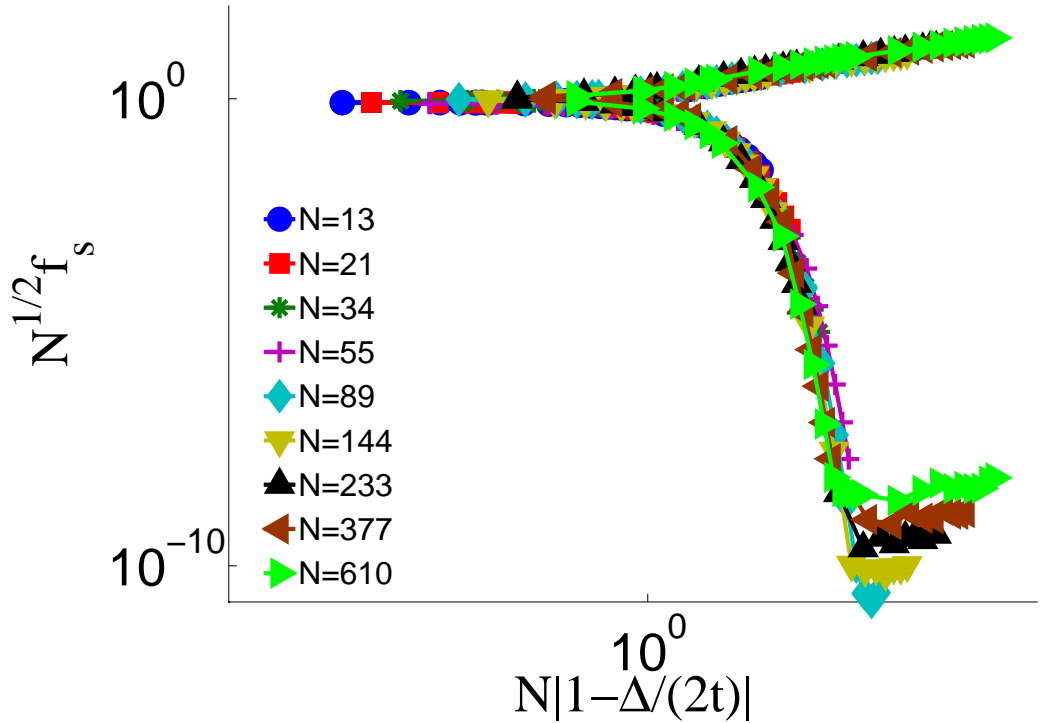}}
  \caption{Double logarithmic scaling plot of the superfluid fraction
    in the case of non-interacting bosons ($\beta=(\sqrt{5}-1)/2 $).
The data for different
    system sizes are collapsing on a single curve once properly
    scaled. Deviations from scaling far
  away from the transition point are visible on the graph, indicating
  that the corresponding points are already outside of the scaling region.}
  \label{fig:fss-noint}
\end{figure}

When approaching the transition from the localised
side, this characteristic length is obviously the localisation
length. On the delocalised state, the identification of the
characteristic length is less straightforward. Using the Aubry-Andr\'e
duality, there is now a dimensionless localisation length $\ell$ 
defined in momentum space,
which measures the number of plane waves of quasi-momentum $k+n\beta$ that
must be combined in the repeated zone scheme to form an extended
wavefunction. When these quasi-momenta are folded back in the first
Brillouin zone, we end up with a wavefunction formed by linear
combination of $\ell$ distinct plane waves, whose momenta are in the
interval $[-\pi/\alpha,\pi/\alpha]$. The typical distance between these different
momenta can be estimated to be of the order of $2\pi/(\ell \alpha)$. Thus, in
order to resolve the difference between these momenta, it is necessary
to be in a system of size $N$ such that $2\pi/(N \alpha) \ll 2\pi/(\ell \alpha)$
i. e. $N\gg \ell$. Thus, the length $\ell$ has the interpretation of
the minimal system length necessary to distinguish a extended state from a
localised state. We thus identify the length $\ell \alpha$ to $\xi(\Delta) \alpha$
in the extended regime. Using the exact result on the localisation
length in the Aubry-Andr\'e model, we have $\xi(\Delta)=1/|\ln
(\Delta/2t)|\sim 1/|\Delta-2t|$. Thus, the exponent $\nu=1$ on both 
sides of the transition.
Using this expression for $\xi(\Delta)$,
we have been able to collapse our numerical data on a single curve, as
can be seen on Fig.~\ref{fig:fss-noint}
using an exponent $a=1/2$. This leads to a superfluid fraction
vanishing as $|2t-\Delta|^{1/2}$ as the transition is approached from
the extended side in the thermodynamic limit.

\subsection{Weakly interacting case}
\begin{figure}
% if it does not compile replace by .eps
\resizebox{0.75\columnwidth}{!}{
\includegraphics{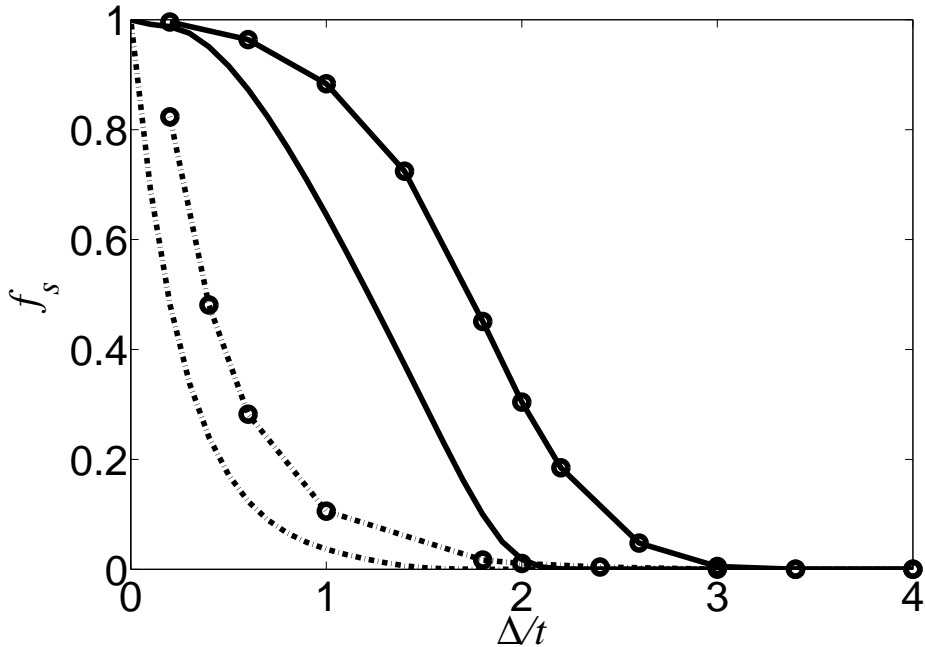}}
\caption{Comparison of the superfluid fraction for a noninteracting
  (straight lines, $U=0$) and weakly interacting Bose gas (lines with circles,
  $U/t=0.02$) in a quasiperiodic potential as a function of
  the height of the secondary lattice $\Delta/t$ at varying choices of
  the periodicity parameter: solid lines, $\beta=(\sqrt{5}-1)/2$;
  dot-dashed lines, $ \beta=1032/862$. Note the enhancement of the
  superfluid fraction by the interaction for both cases.
  For all the curves we have used
  a number of lattice sites $N=50$ and a filling factor $1/2$.}
\label{fig:3}       % Give a unique label
\end{figure}

Using DMRG we can access  the regime of arbitrary interaction
strength \cite{nostro}.  In particular, we consider here the
effect of weak repulsive interactions. In this regime, interaction
reduces localisation, as repulsion keeps particles from coming on
top of each other.  Figure \ref{fig:3} shows the superfluid
fraction as a function of the height of the secondary lattice
$\Delta$. The comparison with the noninteracting case shows that
in the interacting case the localisation transition is shifted to
higher values of the secondary lattice $\Delta$. The delocalised
phase, having both non-vanishing compressibility and non-vanishing
superfluid fraction is in a Luttinger liquid
phase.\cite{nostro,haldane} In particular, it displays a
quasi-long-range superfluid order, with the single particle
density matrix decaying as a power law as a function of distance.
This also gives a power law divergence of the momentum
distribution at low momentum in the infinite-system
limit\cite{nostro}. The observation of the shift of the transition
could provide a sensitive measure of the interaction strength $U$,
hence of the scattering length $a_s$.

\subsubsection{Finite size scaling}
\begin{figure}
  \resizebox{0.95\columnwidth}{!}{
  \includegraphics{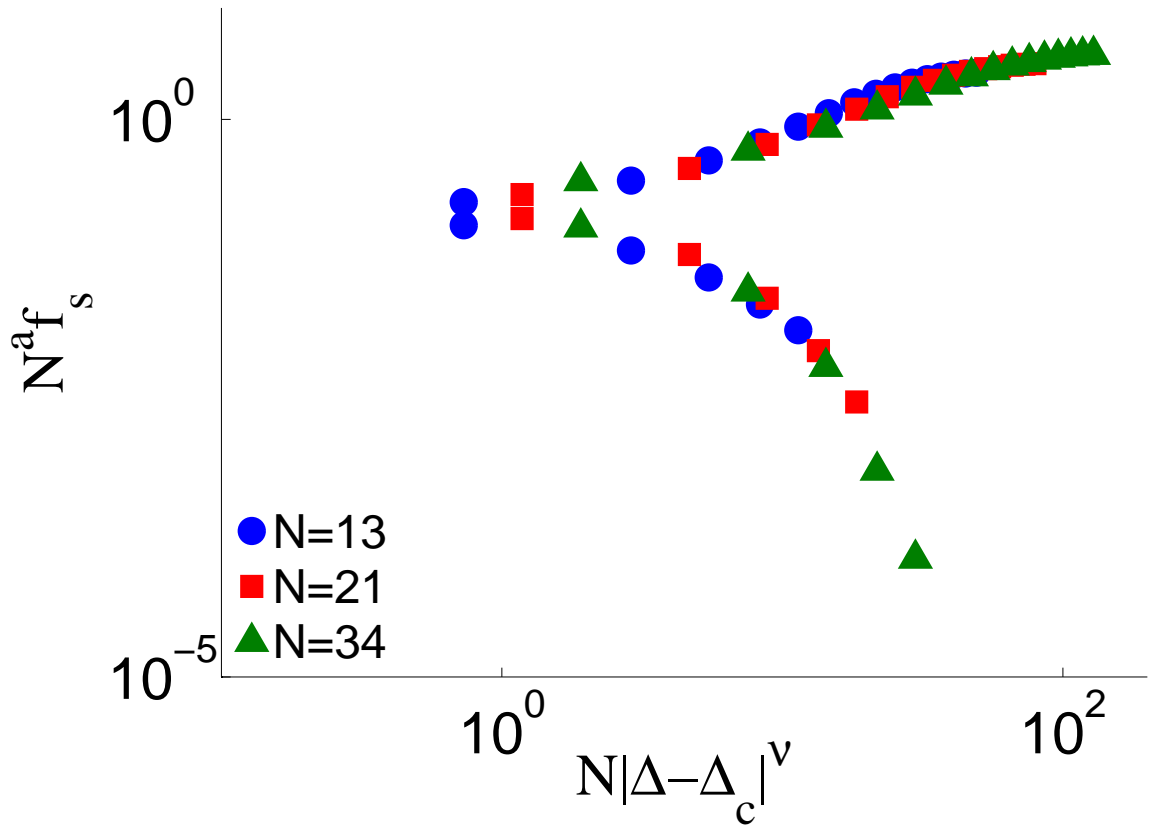}}
  \caption{(Colour online) Double logarithmic scaling plot of the
    superfluid fraction in the case $U/t=0.02$. The parameters are
    $\Delta_c=3.2 t$, $\nu=1.25$, $a=0.38$. }
  \label{fig:fss-interacting}
\end{figure}

In the interacting case, we can also develop a finite size scaling
Ansatz analogous to  (\ref{eq:fss-ansatz}). We now assume that
$\xi(\Delta) \sim |\Delta -\Delta_c(U)|^{-\nu}$, where $\Delta_c(U)$
is the transition point in the presence of interaction $U$. The main
difficulty in applying finite size scaling to the transition in the
interacting case is that in contrast to the non-interacting case of
section~\ref{sec:fss} we do not have \emph{a priori} knowledge of
$\Delta_c(U)$. However, we note that in the localized phase, $f_s$ is decreasing exponentially with the system size, and at the critical point, according to finite-size scaling is decreasing as a power law. By plotting $f_s$ as a function of $\Delta$ with $U=0.02t$, we find that $f_s$ is a non-decreasing function of 
$N$ for $\Delta/t<3.2$. Thus, the localized phase is obtained for $\Delta>3.2t$. 
We have assumed that $\Delta_c=3.2t$ was the location of the critical point,
and fitted the $N$ dependence of $f_s(\Delta_c,N)$ to obtain the exponent $a=0.38$. We have been able to collapse the curves (see  Fig.~\ref{fig:fss-interacting}) using $\nu=1.25$. The
collapse of the data points is a verification of the scaling
hypothesis. 

\section{Momentum distribution}
\label{sec:momdistr}
The momentum distribution is one of the experimentally accessible
observables \cite{roati08}, and allows to distinguish among the
extended and localised regimes.  We have calculated the momentum
distribution for various choices of the quasiperiodicity parameter
$\beta$, both in the noninteracting and in the weakly interacting
case.

\begin{figure}
% if it does not compile replace by .eps
\resizebox{0.95\columnwidth}{!}{
\includegraphics{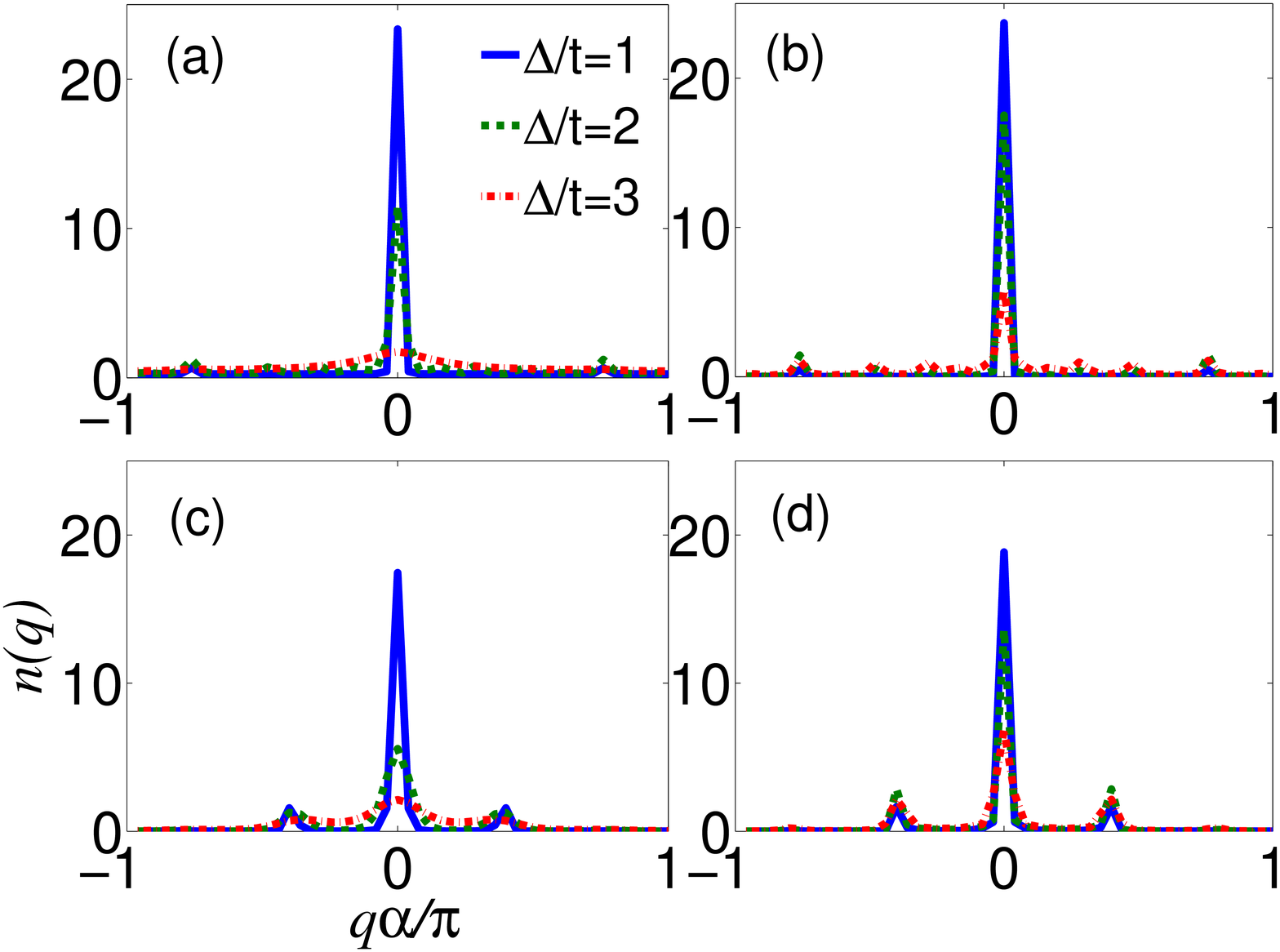}}
\caption{(Colour online) Momentum distribution (in units of the primary lattice spacing $\alpha$) of a Bose gas in the quasiperiodic lattice, for various choices of
the
 height of the secondary lattice $\Delta/t=1$ (solid lines), $\Delta/t=2$ (dashed lines) and $\Delta/t=3$ (dot-dashed lines); of the interaction strength $U/t=0$ (left panels) and $U/t=0.02$ (right panels) and of the periodicity parameter  $\beta=(\sqrt{5}-1)/2$ (top panels) and $ 1032/862$ (bottom panels). For all the curves we have used a number of lattice sites $N=50$ and a filling factor $1/2$.}
\label{fig:4}       % Give a unique label
\end{figure}

Our results are reported in Fig. \ref{fig:4}. In the extended
regime, in addition to the central peak at $q=0$ side peaks in the
momentum distribution appear, at a position related to the value
of the periodicity of the secondary lattice and corresponding to
the beating of the two lattices. The position of the secondary
peaks is at $Q=\pm \frac{2\pi}{\alpha}(1-\beta)$. In the localised
regime, the central peak of the momentum distribution is broadened
till becomes of the order of the Brillouin zone (Fig. \ref{fig:4}
(a) and (b))), while the secondary peaks are strongly suppressed.
This is in agreement the experimental findings\cite{roati08}. The
broadening of the distribution in momentum space indicates that
the states in real space are more and more localised.  As
interactions are turned on, we see a restoration of the peaks of
the momentum distribution as a signature of the enhanced
superfluid behaviour found by the analysis of the superfluid
fraction.

\begin{figure}
% if it does not compile replace by .eps
\resizebox{0.99\columnwidth}{!}{
\includegraphics{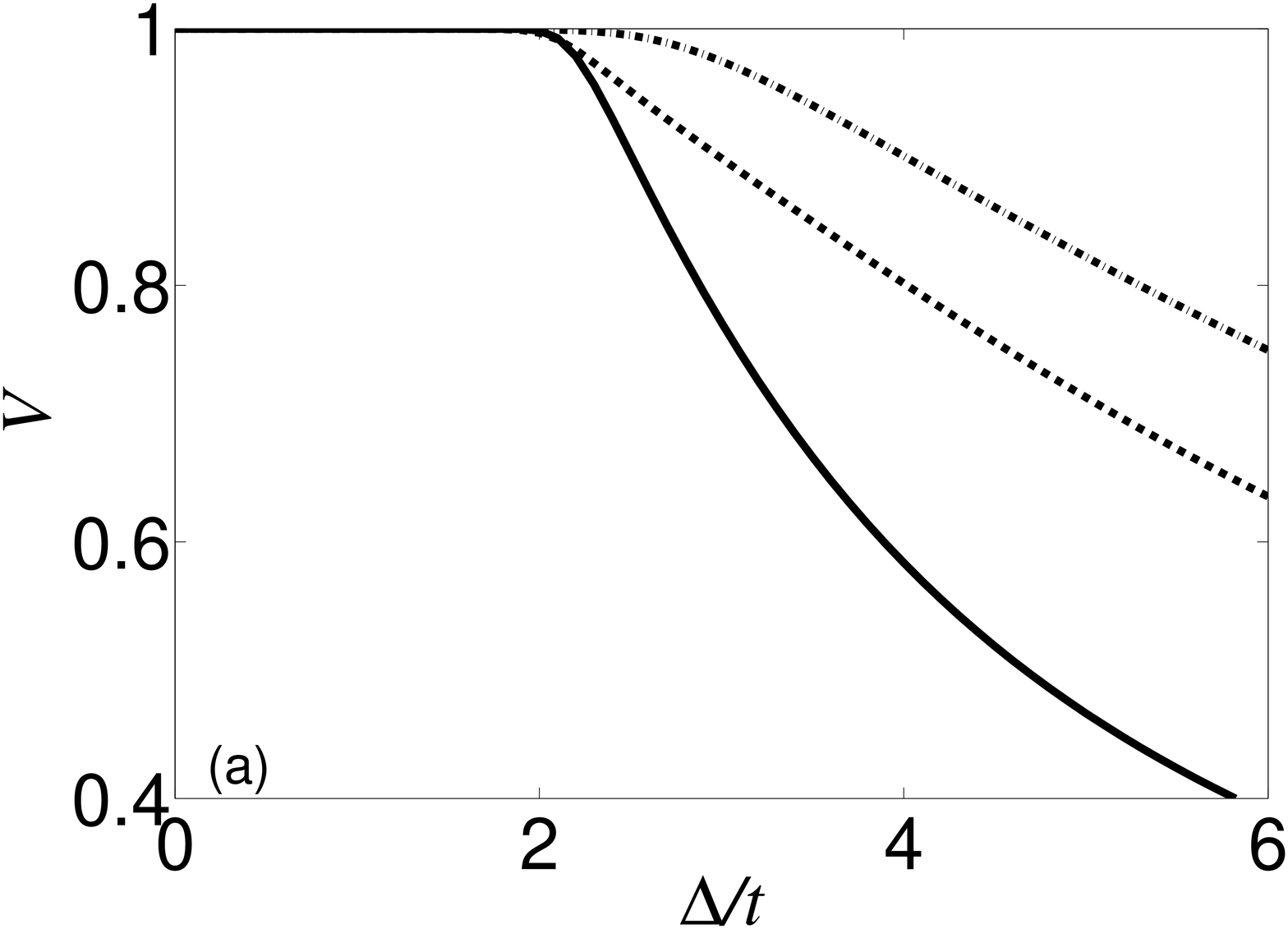}
\includegraphics{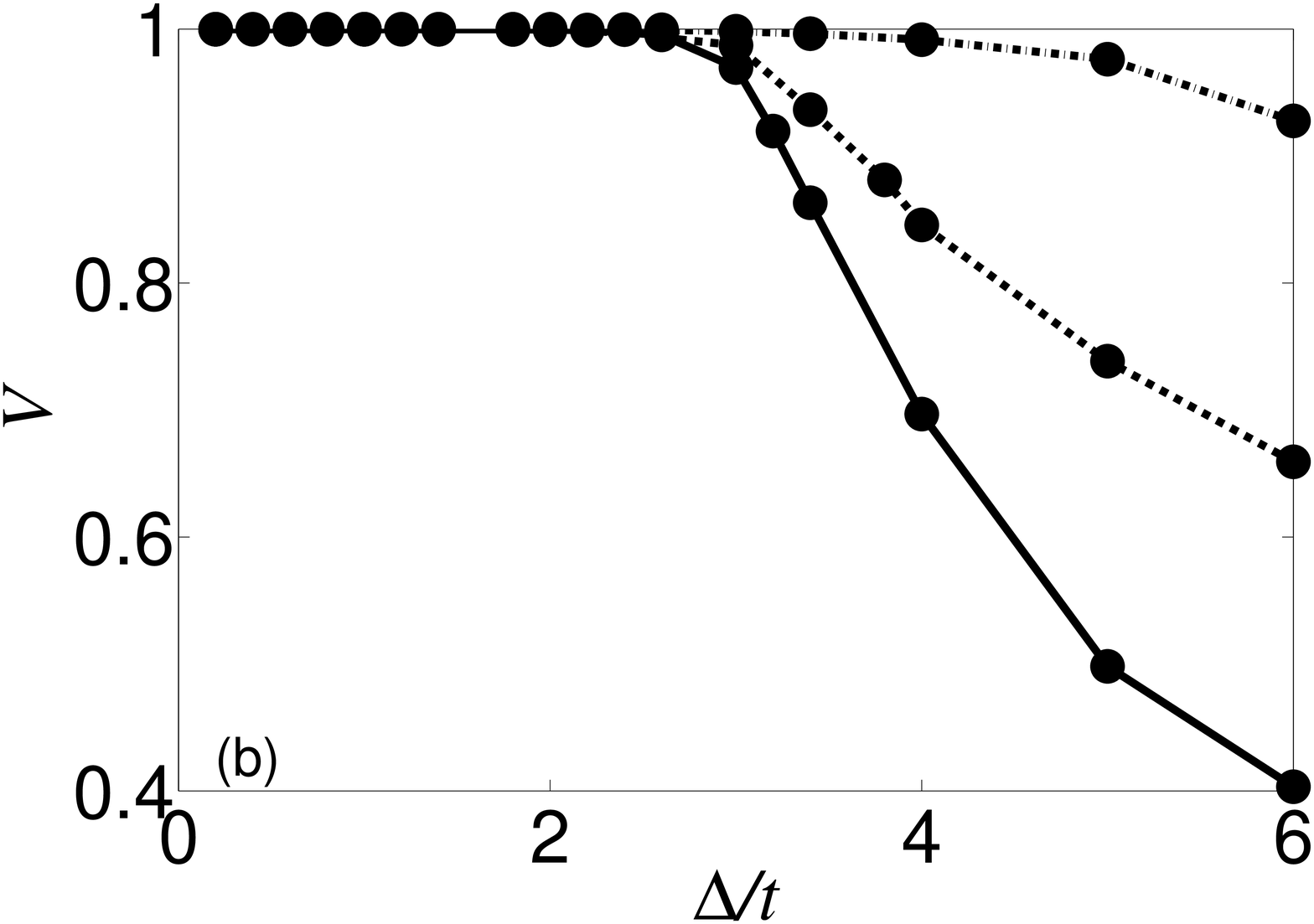}
}
\caption{Visibility of the momentum distribution (adimensional) of a
  Bose gas in the quasiperiodic lattice, as a function of the height
  of the secondary lattice $\Delta/t$ for three choices of the
  quasiperiodicity parameter $\beta$  (solid line:
  $\beta=(\sqrt{5}-1)/2$, dotted line $\beta=830/1076$, dash-dotted
  line $\beta=1032/862$); left panel, noninteracting case; right panel interacting case with $U=0.02t$.  For all the curves we have used a number of lattice sites $N=50$ and a filling factor $1/2$.}
\label{fig:5}       % Give a unique label
\end{figure}

The transition from extended to localised phase has been characterised
in the experiment \cite{roati08} by studying the visibility of the
secondary peaks in the momentum distribution.  In Fig. \ref{fig:5} we
show the visibility of the momentum distribution, defined as
$V=[n(0)-n(k_1)]/[n(0)+n(k_1)]$, both for the noninteracting and for the interacting case.
 The loss of visibility starts to become
important at values of the height of the secondary lattice $\Delta/t$
which depend on the choice of $\beta$ and of the interaction strength.
In the case of an irrational $\beta=(\sqrt{5}-1)/2$ and in the noninteracting limit the visibility starts to decrease right at the transition point  $\Delta/t=2$, while for other choices of $\beta$ as in the experiments the visibility remains close to one to larger values of $\Delta/t$, indicating a smoother crossover from extended to localised states. Similarly, in the interacting case the visibility starts to decrease from its value one at larger value of $\Delta/t$, reflecting the shift of the transition point.

\section{Summary and concluding remarks}
\label{sec:summary}

In summary, we have analysed the superfluid to
Anderson-localisation transition for an interacting Bose gas
subjected to a quasiperiodic potential, focussing on the case of a
finite-size lattice. We have investigated the effects of changing
the values of the quasiperiodicity parameter $\beta$ and of the
interactions. In the case of periodic boundary conditions, by
studying the superfluid fraction we have found that  the
transition is sharper for irrational values of $\beta$ and
smoother when several minima in the potential have almost the same
energy. We have also found that the transition is shifted to
higher values of the secondary lattice height when the repulsive
interactions are taken into account. Implementation of this idea  could lead to
an accurate method to measure  of the interaction coupling
strength, which is directly related to the
s-wave scattering length . In the case of open boundary
conditions, the inverse participation ratio is introduced as a
good indicator of the transition, and its scaling analysis
reproduces again the change of character of the transition from
irrational values of $\beta$ to rational ones (with a large
denominator). We have furthermore shown that also in the
interacting case the change of the ground state of the system
might be probed by analysing the momentum distribution of the
gas, which displays side peaks in the superfluid phase but not
 in the deeply localised phase,  and the visibility of the momentum
distribution. In perspective, a detailed analysis of the shift of
the transition at increasing the interaction strength, as well as
the inclusion of a harmonically trapping potential\cite{note1},
seem to us valuable extensions of the present work.

\section*{Acknowledgements}
R.C.
acknowledges funding from Marie-Curie Intra-European fellowship,
X.D. funding from CNRS and A.M. from the MIDAS-STREP project.


\begin{thebibliography}{}
\bibitem{oberthaler06} O. Morsch and M. Oberthaler, Rev. Mod. Phys. {\bf 78}, (2006)
179
\bibitem{feynman82} R. P. Feynman, Int. J. Theor.
Phys. {\bf 21}, (1982) 467
\bibitem{Anderson}  P. W. Anderson, Phys. Rev.
{\bf 109}, (1958) 1492.
\bibitem{loc_wave_1}  M. P.  Van Albada and A. Lagendijk, Phys. Rev.
Lett. {\bf 55}, (1985)  2692
\bibitem{loc_wave_2} D.S. Wiersma, P. Bartolini, A. Lagendijk, and
  R. Righini,  Nature {\bf 390}, (1997) 671
\bibitem{loc_wave_3}  M.  St\"orzer, P.  Gross, C.M.  Aegerter  and
 G. Maret Phys. Rev. Lett. \textbf{96} (2006) 063904
\bibitem{HefeiHu2008} Hefei Hu, A. Strybulevich, J.H. Page, S.E. Skipetrov and B.A. van Tiggelen, Nature Phys. \textbf{4} (2008) 945
\bibitem{lee} P. A. Lee and T. V. Ramakhrishnan
  Rev. Mod. Phys. \textbf{57} (1985) 287
\bibitem{Albergamo} F. Albergamo, J. Bossy,  J. V. Pearce, H. Schober
  and H. R. Glyde Phys. Rev. B 76 (2007), 064503
\bibitem{Oudenaarden} A. van Oudenaarden,  S . J. K. V{\'a}rdy and
 J.E. Mooij Phys. Rev. Lett. \textbf{77} (1996) 4257
\bibitem{zoller_dis} B. Damski, J. Zakrzewski, L. Santos,
P. Zoller, and M. Lewenstein, Phys. Rev. Lett. {\bf 91},  (2003)
080403
\bibitem{inguscio_05}J.E.  Lye, L. Fallani,
M. Modugno, D. Wiersma,  C. Fort, and M. Inguscio, Phys. Rev.
Lett. {\bf 95}, (2005) 070401

\bibitem{inguscio_07}L. Fallani, J.E.  Lye,
V. Guarrera, C. Fort, and M. Inguscio,  Phys. Rev. Lett. {\bf
98}, (2007) 130404

\bibitem{aspect08} J. Billy, V. Josse, Z. Zuo, A. Bernard,
  B. Hambrecht, P. Lugan, D. Cl\'ement, L. Sanchez-Palencia, P. Bouyer
  and A. Aspect, Nature \textbf{453}, (2008) 891

\bibitem{roati08} G. Roati, C. D'Errico, L. Fallani, M. Fattori, C. Fort, M. Zaccanti,
G. Modugno, M. Modugno and M. Inguscio, Nature {\bf 453}, (2008) 895

\bibitem{diener01} R. B. Diener, G. A. Georgakis, J. Zhong, M. Raizen, and Q. Niu, Phys. Rev. A {\bf 64},  (2001) 033416
\bibitem{grempel82} D. R. Grempel, S. Fishman, and R. E. Prange,
  Phys. Rev. Lett. {\bf 49}, (1982)
833
\bibitem{sokoloff85} J. B. Sokoloff Phys. Rep. \textbf{126} (1985) 189
\bibitem{harper} P.G. Harper, Proc. Phys. Soc. London A {\bf 68},   (1955) 874
\bibitem{aubry} S. Aubry, in {\it Solitons and Condensed Matter Physics},
edited by A. R. Bishop and T. Schneider (Springer, New York,
1978); S. Aubry and G. Andre, Ann. Isr. Phys. Soc. {\bf 3},  (1980) 133;
 S. Aubry, J. Phys. (Paris) {\bf 44},  (1983) 147
\bibitem{hofstadter} D.R. Hofstadter, Phys. Rev; B {\bf 14} (1976) 2239
\bibitem{almostMathieu} Svetlana Ya. Jitomirskaya, Ann. Math. {\bf
    150},  (1999) 1159
\bibitem{Twose} N. F. Mott and W. D. Twose, Adv. Phys. \textbf{10}
  (1961) 107.

\bibitem{schulte_0506} T. Schulte, S. Drenkelforth, J. Kruse, W. Ertmer, J. Arlt, K.
Sacha, J. Zakrzewski, and M. Lewenstein, Phys. Rev. Lett. {\bf
95},  (2005) 170411; T. Schulte, S. Drenkelforth, J. Kruse, R.
Tiemeyer, K. Sacha, J. Zakrzewski, M. Lewenstein, W. Ertmer, and
J. J. Arlt, New J. Phys. {\bf 8},   (2006) 230
\bibitem{Roscilde08a} T. Roscilde, Phys. Rev. A, {\bf 77}  (2008) 063605
\bibitem{Roscilde08b}  T Roscilde, arXiv:0804.2769
\bibitem{Fisher} M.P. Fisher, P.B. Weichman, G. Grinstein, and D.S. Fisher, Phys. Rev. B
{\bf 40} (1989) 546.
\bibitem{GiaSchu} T. Giamarchi and H. J. Schulz, Phys. Rev. B, {\bf
    37} (1988) 325
\bibitem{Doty} C. Doty and D. S. Fisher, Phys. Rev. B \textbf{45} (1992),
  2167
\bibitem{Runge} K. Runge and G.T. Zimanyi Phys. Rev. B, \textbf{49}
  (1994), 15212.
\bibitem{Batrouni} G.G. Batrouni and R.T. Scalettar, Phys. Rev. B, {\bf 46} (1992) 9051
\bibitem{Svistunov} B.V. Svistunov,  Phys. Rev. B, {\bf 54} (1996) 16131,
 N. V. Prokof'ev and B. V. Svistunov,  Phys. Rev. Lett., {\bf 80} (1998) 4355
 \bibitem{monien96} J. K. Freericks and H. Monien Phys. Rev. B {\bf
  53},  (1996) 2691
\bibitem{DMRG_disorder} P. Schmitteckert, T. Schulze, C. Schuster,
P. Schwab and U. Eckern,  Phys. Rev. Lett., {\bf 80} (1998) 560
\bibitem{Roux08} G. Roux, T. Barthel, I. P. McCulloch, C. Kollath, U. Schollwoeck, and T. Giamarchi,  Phys. Rev. A, {\bf 78}(2008) 023628
\bibitem{nostro} Xialong Deng, R. Citro, A. Minguzzi, and E. Orignac, Phys. Rev. A, {\bf 78} (2008) 013625
\bibitem{jaksch} D. Jaksch et al., Phys. Rev. Lett., {\bf 81}(1998) 3108
\bibitem{Buchler03} H.P. Buechler, G. Blatter and W. Zwerger, Phys. Rev. Lett., {\bf 90}(2003) 130401
\bibitem{bloch_review_08} I. Bloch, J. Dalibard and W. Zwerger,
Rev. Mod. Phys. {\bf 50}, (2008) 885
\bibitem{inguscio_07_pra} J.E. Lye, L. Fallani, C. Fort, V. Guarrera, M. Modugno, D.S.
Wiersma, and M. Inguscio, Phys. Rev. A {\bf 75}, (2007) 061603(R).
\bibitem{thouless74} D. J. Thouless, Phys. Rep. \textbf{13} (1974) 93
\bibitem{White_DMRG} S.R. White, Phys. Rev. Lett. {\bf 69}, (1992) 2863; Phys. Rev. B {\bf 48}, (1993) 10345
\bibitem{DMRG_review} U. Schollwoeck, Rev. Mod. Phys. {\bf 77}, (2005)
  259; R.M. Noack and S. Manmara AIP Conf. Proc. {\bf 789}, (2005)
  93; K. Hallberg Adv. Phys. \textbf{55}, (2006) 477.

\bibitem{barber} M. N. Barber in \textit{Phase Transitions and
    Critical Phenomena} vol. 8, edited by C. Domb and J. L. Lebowitz
  (Academic Press, New York, 1983).
\bibitem{note1} The presence of a shallow trapping potential would
not alter the conclusions of our analysis.
\bibitem{haldane} F. D. M. Haldane, Phys. Rev. Lett. {\bf 81}, (1981)
  1840.
\end{thebibliography}
\end{document}